\documentclass{iau307}
\usepackage{graphicx}
\usepackage{natbib}
\usepackage{url}
\usepackage{dtklogos}
\bibpunct{(}{)}{;}{a}{}{,}

\title[Linear Spectropolarimetry to measure wind asymmetry] 
{Linear line spectropolarimetry as a new window to measure 2D and 3D wind geometries}

\author[Jorick S. Vink]   
{Jorick S. Vink$^1$}

\affiliation{$^1$Armagh Observatory, College Hill, BT61 9DG Armagh, Northern Ireland \\ email: {\tt jsv@arm.ac.uk}}

\pubyear{2014}
\volume{307} 
\pagerange{}
\jname{New windows on massive stars: asteroseismology, interferometry, and spectropolarimetry}
\editors{G. Meynet, C. Georgy, J.H. Groh \& Ph. Stee, eds.}
\begin{document}

\maketitle

\begin{abstract}
Various theories have been proposed to predict how mass loss depends on the stellar rotation 
rate, both in terms of its strength, as well as its latitudinal dependence, crucial for our 
understanding of angular momentum evolution. Here we discuss the tool of linear 
spectropolarimetry that can probe the difference between mass loss from the pole versus the 
equator. Our results involve several groups of O stars and Wolf-Rayet stars, involving Oe stars, Of?p stars, Onfp stars, as well as the 
best candidate gamma-ray burst progenitors identified to date.
\keywords{techniques: polarimetric, circumstellar matter, stars: early-type, stars: emission-line, Be, stars: mass loss, stars: rotation, stars: winds, outflows, stars: Wolf-Rayet}
\end{abstract}

\firstsection % if your document starts with a section,
\section{Introduction}

\noindent
Ultimately, we would like to understand massive stars and their progeny
both locally as well as in the distant Universe. What is clear is 
that rotation, mass loss, and the link between them, play a pivotal role 
in the fate of massive stars. 
However, in order to test mass-loss predictions for rotating stars, we 
need to probe the density contrast between the stellar pole and 
equator. In the local Universe, this may potentially be achievable through 
the technique of long-baseline interferometry, as discussed during this 
meeting. However, in order to determine wind asymmetry in the more 
distant Universe we necessarily rely on the technique of {\it linear} 
spectropolarimetry. The only limiting factor is then the collecting 
power of the mirror of the largest  
telescopes.

\section{2D Wind Predictions}

\noindent
Until 3D radiation transfer models with 3D 
hydrodynamics become available, theorists 
have necessarily been forced to make assumptions with respect
to either the radiative transfer (e.g. by assuming a power law
approximation for the line force due to \citealt{CAK}) 
or the hydrodynamics, e.g. by assuming an empirically motivated 
wind terminal velocity in Monte Carlo predictions \citep{AL85,V00}. Albeit 
recent 1D and 2D models of \cite{MV08,MV14} no longer 
require the assumption of an empirical terminal wind velocity. 

There are 2D wind models on the market that 
predict the wind mass loss predominately emanating
from the equator \citep{FA86,BC93,LP91,Pel00}, whilst other models predict 
higher mass-loss rates from the pole, in particular as a result of 
the \cite{vonZeipel24} theorem, resulting in a larger polar Eddington factor 
than the equatorial Eddington factor \citep{O96,PP00,MM00,MV14}. 

The key point is that mass loss from the equator results in more angular 
momentum loss than would 1D spherical or 2D polar mass loss, so we need 
2D data to test this.

\section{Line polarization versus depolarization}

\begin{figure}
\begin{center}
\includegraphics[width=\textwidth]{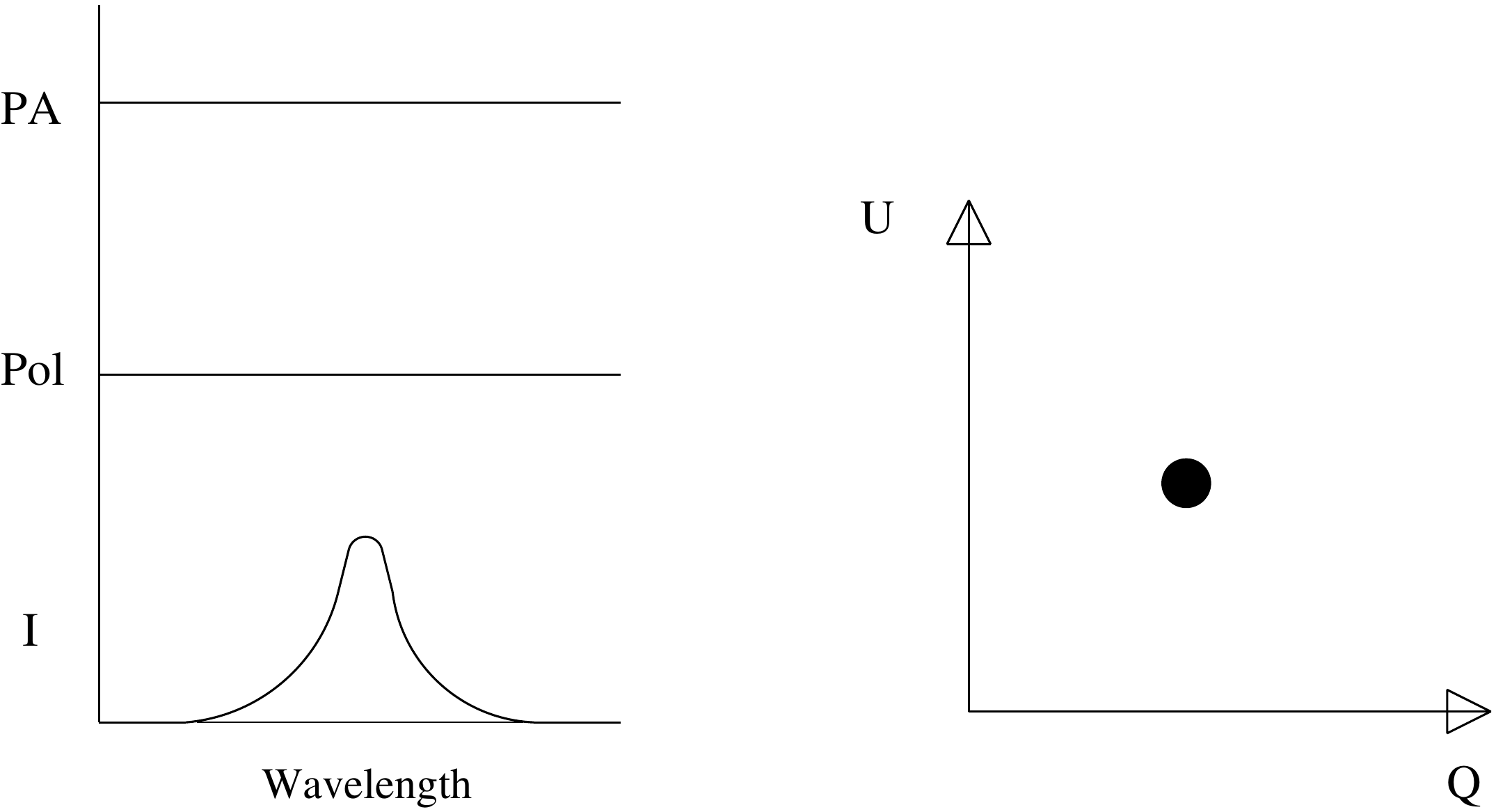}
\caption{Cartoon indicating ``no line effect'. On the left, 
polarization spectrum ``triplot'' and a Stokes $QU$ diagram on the right.
A typical Stokes I emission is shown in the lower panel of the triplot, 
the \%Pol in the middle panel, while the Position Angle (PA) is 
sketched in the upper panel of the triplot. See \citet{V02} for further details.}
\label{fig1}
\end{center}
\end{figure}

\noindent
Whilst {\it circular} Stokes $V$ spectropolarimetry is oftentimes employed 
to measure stellar magnetic fields, {\it linear} Stokes $QU$ polarimetry can 
be utilized to measure large-scale 2D asymmetry in a stellar wind or any other 
type of circumstellar medium, such as a disk. In this sense, the Stokes $QU$
plane plays an analogous role to the interferometric $UV$ plane, with the 
additional advantage that it can measure the smallest spatial scales, such
as the inner disk holes of order just a few stellar radii 
in pre-main sequence (PMS) stars \citep{V05}, which would otherwise
remain ``hidden'', or the driving region of stellar winds in massive stars, that we explore in the following.

In principle, linear continuum polarimetry would already be able to inform us about the presence 
of an asymmetric (e.g. a disk or flattened wind) structure on the sky, but in practice, this 
issue is complicated by the roles of intervening circumstellar and/or interstellar 
dust, as well as instrumental polarization. The is one of the reasons linear {\it spectro}polarimetry, measuring
the change in the degree of linear polarization across emission lines is such a powerful tool, as ``clean'' or ``intrinsic'' information
can be directly obtained from the $QU$ plane. 
The second reason is the additional bonus that it may provide kinematic information
of the flows around PMS as well as massive stars. 

\begin{figure}
\begin{center}
\includegraphics[width=\textwidth]{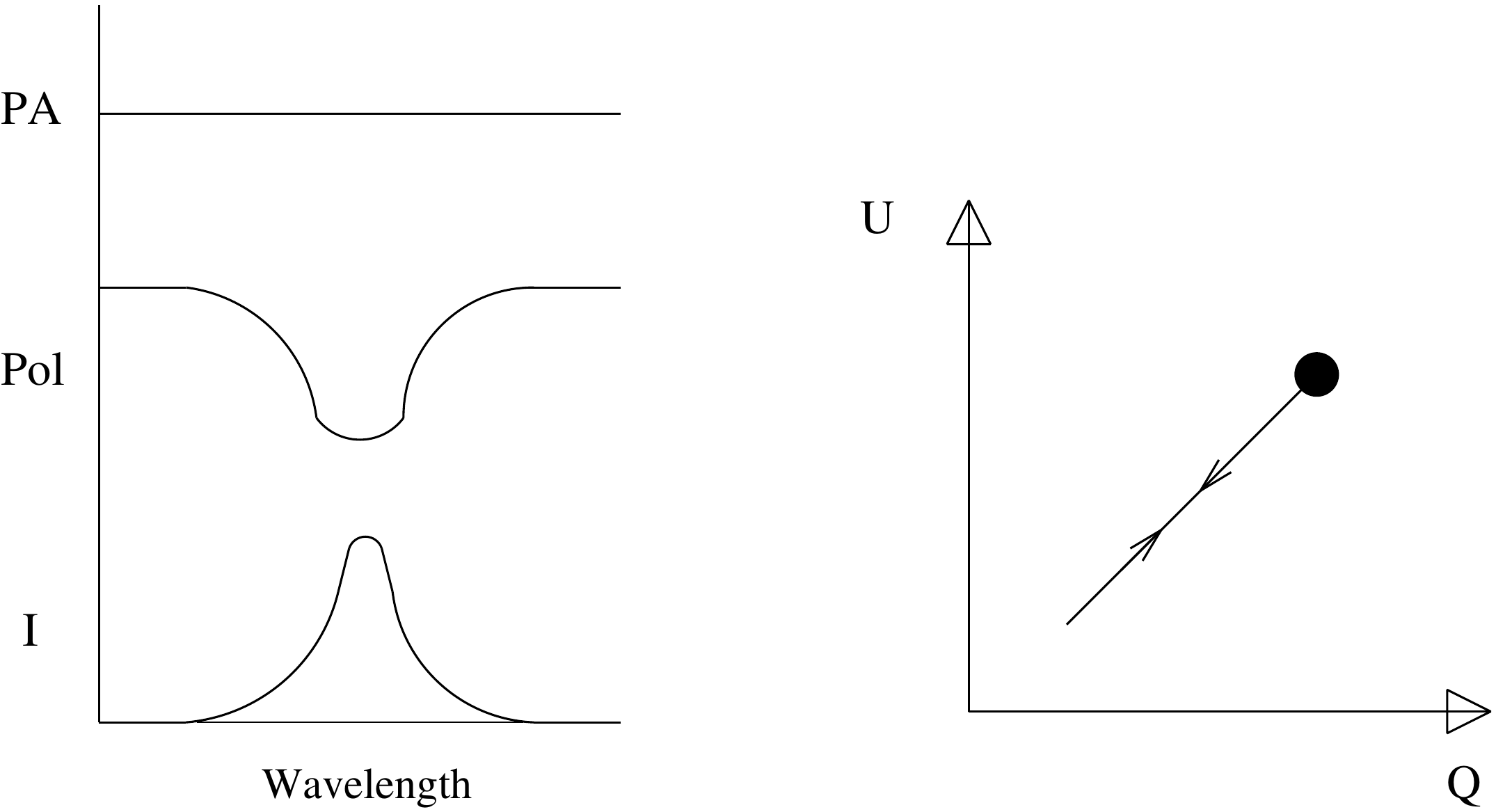}
\caption{Cartoon indication ``depolarization'' or ``dilution''. Note that the 
depolarisation across the line is as broad as the Stokes $I$ emission. Depolarisation
translates into Stokes $QU$ space as a linear excursion. See \citet{V02} for further details.}
\label{fig2}
\end{center}
\end{figure}

Figures 1-3 show linear line polarization 
cartoons (both in terms of polarization ``triplot'' spectra and Stokes $QU$ planes) 
for the case that the spatially unresolved object under consideration is (i) 
spherically symmetric on the sky showing ``no line effect'', (ii) asymmetric  
showing line ``depolarization'' where the emission line simply acts to ``dilute'' the polarized 
continuum, or (iii) cases where the line effects are more subtle, involving 
position angle (PA) flips across intrinsically polarized lines.

\begin{figure}
\begin{center}
\includegraphics[width=\textwidth]{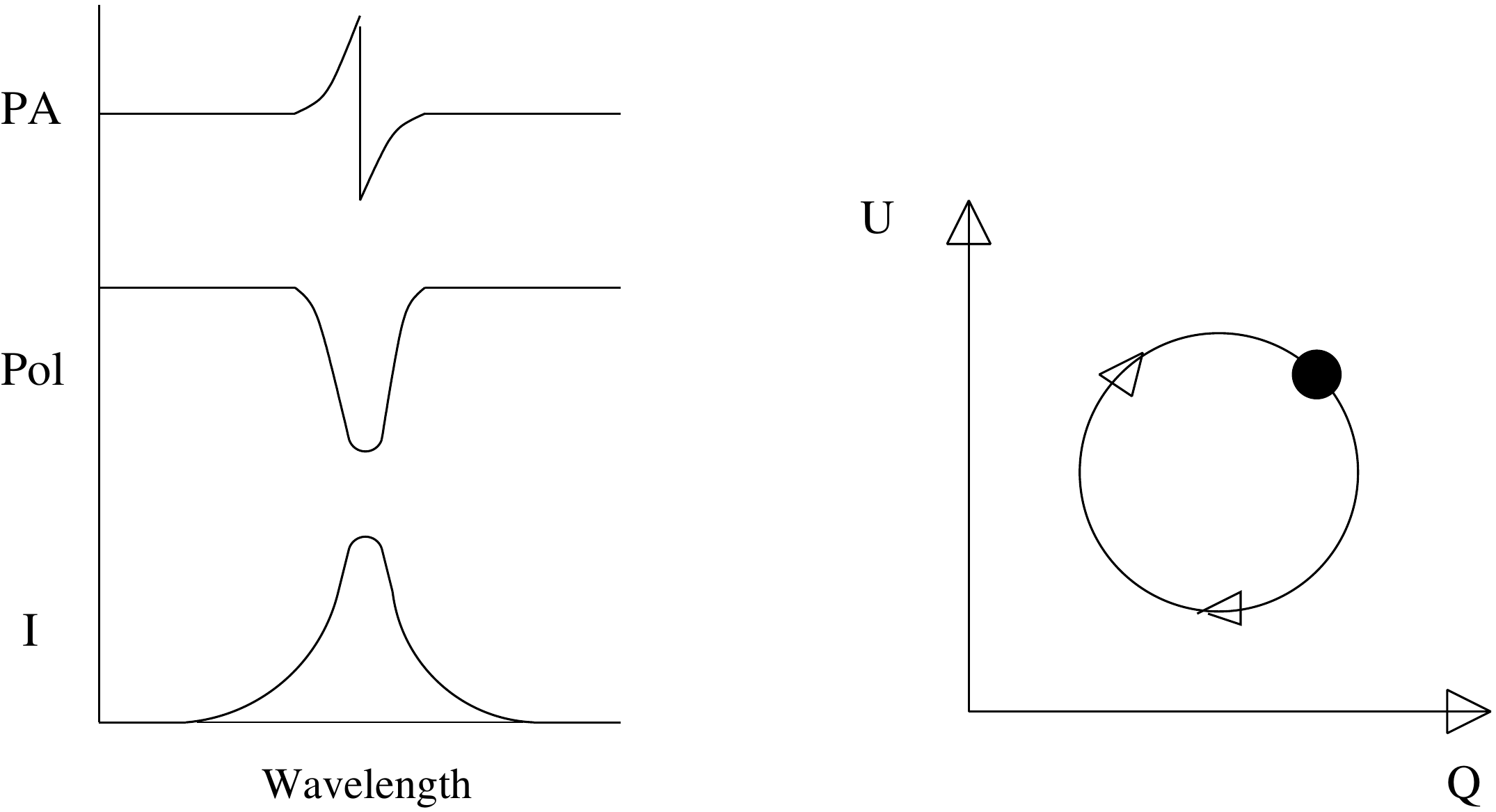}
\caption{Cartoon indicating a compact source of line photons scattered off a {\it rotating} disk. 
Note that the polarisation signatures are relatively narrow compared 
to the Stokes $I$ emission. The PA flip is associated with a loop
in Stokes $QU$ space. See \citet{V02,V05} for further details.}
\label{fig3}
\end{center}
\end{figure}

\begin{figure}
\begin{center}
\includegraphics[width=0.48\textwidth]{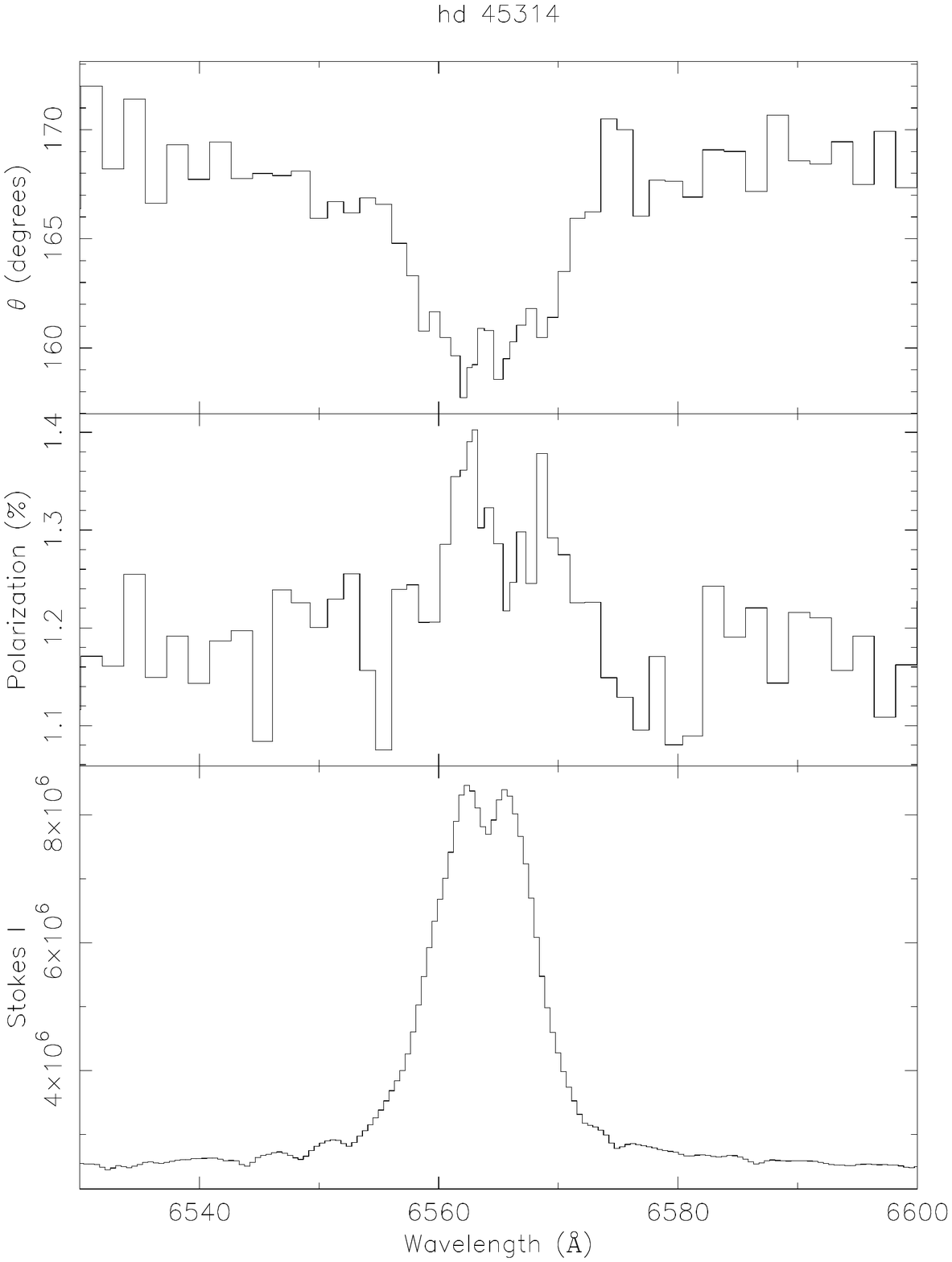}
\includegraphics[width=0.48\textwidth]{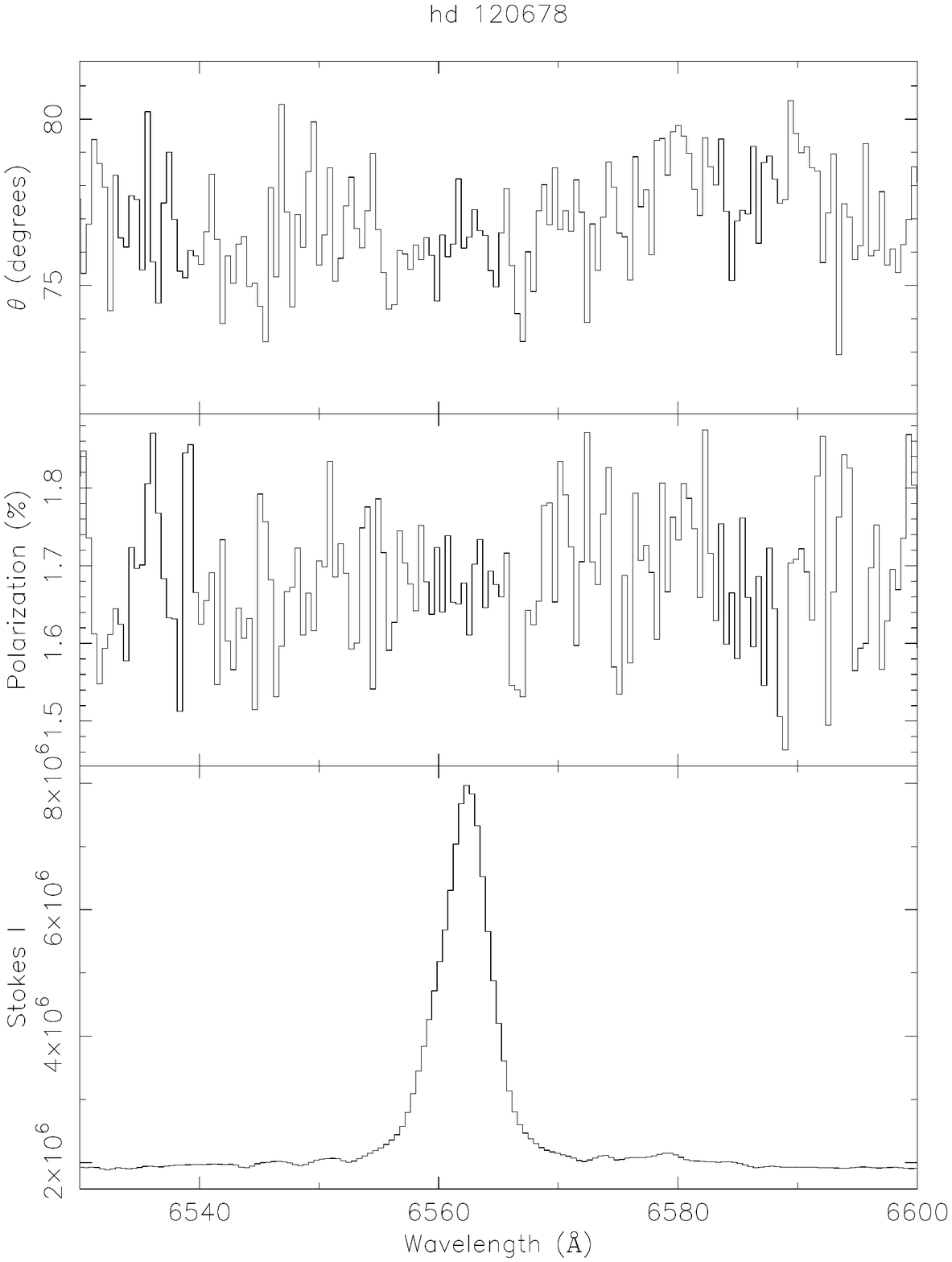}
\caption{H$\alpha$ line polarization ``triplots'' of the Oe stars HD\,45314 and HD\,120678. 
%The data are rebinned such that the 1$\sigma$ error in the polarisation
%corresponds to 0.05\% as calculated from photon statistics. 
HD\,45314 shows a line effect indicating that it 
is intrinsically polarized, but  HD\,120678 is not intrinsically polarized. See \citet{V09} for further details.}
\label{fig4}
\end{center}
\end{figure}

Whilst the third situation of intrinsic line polarization in a rotating disk has been 
encountered in PMS (see \citealt{V05}), it is the second case of ``depolarization'' that
is most familiar to the massive-star community through its application to classical 
Be stars, starting as early as the 1970s (see the 
various works by Poeckert, Marlborough, Brown, Clarke, and McLean). 
Interestingly, the same method has in more recent years also been applied to 
Oe stars, the alleged more massive counterparts of Be stars, see Fig.\,4. 
Note that although the Oe star HD\,120678 (on the right hand side of Fig.\,4) has a 
significant observed level of linear polarization, the 
lack of a line effect implies that the object is not intrinsically polarized \citep{V09}. 
This could either mean the object is spherically symmetric or that it has a disk that is 
too ``pole on'' to provide intrinsic polarization. It is for these reasons vital
to consider a {\it sample} of objects. For Oe stars \cite {V09} found that the incidence of line 
effects (1/6) was much lower than for Be stars. This implies 
that the chance the Oe and the Be stars are drawn from the same parent distribution 
is small, providing relevant constraints on the formation of Be stars. 

\section{Survey results of O and Wolf-Rayet winds}

\noindent
We now turn to more massive stars with stronger winds than Oe/Be stars. 
Linear spectropolarimetry results have been performed on relatively large 
samples (of order 40-100) for both O \citep{H02,V09} and Wolf-Rayet (WR) stars 
\citep{H98,V07}, and the key result from these surveys is that the vast
majority of 80\% of them is to first order spherically symmetric. 
This is of key importance for the accuracy of mass-loss predictions from 1D models 
for rotating stars.

However, the above studies also found a number of interesting {\it exceptions}. With respect to O stars, 
\citet{V09} found that certain O-type subgroups involving Of?p and Onfp 
class are more likely polarized than the garden-variety of spherical O-stars. 
For instance, \citet{V09} highlighted that HD\,108 is linearly polarized, which 
may be related to its probably magnetic properties. Indeed, it was later found that 
HD\,108 and several other Of?p stars form a magnetic sub-class. 
The line effects in the Onfp stars (involving famous objects like $\lambda$\,Cep and 
$\zeta$\,Pup) may may involve intrinsic line polarization effects due to the 
rapid rotation of this O-type subgroup in addition to (or instead of) depolarization. 

Turning to WR stars, \citet{V11} and \citet{G12} uncovered 
that the small 20\% minority of WR stars that display a depolarization line effect indicating 
stellar rotation are highly significantly correlated with the subset of WR stars that have ejecta nebulae. 
These objects have most likely only recently transitioned from a red sugergiant (RSG) or luminous blue variable (LBV) 
phase. As these presumably youthful WR stars have yet to spin-down, they are the best 
candidate gamma-ray burst (GRB) progenitors identified to date. However, in our own Milky Way these 
WR stars are still expected to spin down before explosion (due to WR winds). 
However, in lower metallicity ($Z$) environments WR stars are thought to be weaker and 
WR stars in low $Z$ environments, such as those studied in the 
Magellanic Clouds may offer the best way to directly pinpoint 
GRB progenitors \citep{V07}.

\section{Future}

\noindent
In addition to the quest for WR GRB progenitors, there is a whole range of interesting wind physics 
to be constrained from linear spectropolarimetry. The main limitation at this point 
is still sensitivity. We are currently living in an exciting time as we are 
at a point where the possibility of extremely large telescopes (ELTs) may become reality. If these telescopes
materialize with the required polarization optics, we might -- for the first time in history -- 
be able to obtain spectropolarimetric data at a level of precision that has been feasible 
with 1D Stokes $I$ data for more than a century. 
It is really important to note that current 3D Monte Carlo radiative transfer 
is well able to do the required modelling, but the main limitation is the necessary 3D data! 

Another interesting future application will involve polarimetric monitoring. 
Whilst we now know that on large scales the 1D approximation is appropriate 
for stellar winds, we have also become aware of the intrinsic 3D clumpy nature of stellar winds 
on smaller scales (but with macroscopic implications!). In particular the existence of wind 
clumps on small spatial scales near the stellar photosphere \citep{Cant09} has 
been confirmed by linear polarization variability studies \citep{D07}, but to probe 
further -- mapping wind clumps in detail -- we need good monitoring data.

\bibliographystyle{iau307}
\bibliography{vink_iau307}

\end{document}